\documentclass[10pt,a4paper,english]{article}
\usepackage[T1]{fontenc}
\usepackage[latin1]{inputenc}
\usepackage{graphicx}

\makeatletter

\usepackage{a4wide}

\usepackage{babel}
\makeatother
\begin{document}

\title{Fermionic zero modes on a toroidal cosmic string}

\author{Abhijit B. Gadde and Urjit A. Yajnik\\
\textsl{Indian Institute of Technology Bombay, Mumbai 400076}}

\date{}

\maketitle
\begin{abstract}
We consider a toroidal configuration of cosmic string in $3+1$ dimensions
in an abelian Higgs model, a compactification of the Nielsen-Olesen
string. This object is classically unstable. We explicitly compute
the number of permitted zero modes for majorana fermions coupled to 
such a string. As in the case of indefinitely long strings, there 
are $|n|$ zero modes for winding number sector $n$, and correspondingly,
induced fermionic charge $n/2$ which canbe fractional. According to a 
previously proved result, this implies quantum mechanical stability
for objects with odd winding number. The result is of significance
to cosmology in classes of unified theories permitting such cosmic
strings. 
\end{abstract}

\section{Introduction}

Solitons are classically stable solutions of field equations, made
possible when the spontaneous breaking of gauge symmetry premits topologically
non-trivial boundary conditions. Examples of such solitons exist in
one dimension \cite{Jackiw1976} called kinks, where the gauge field
is absent, in two dimensions \cite{Nielsen1973} called Nielsen-Olesen
strings which occur in abelian gauge theories, in three dimensions
\cite{Hooft1974,Polyakov1974}, being `t Hooft-Polyakov monopoles
occuring in Yang-Mills theories. More interesting systems are those
in which the interaction of the fermi fields with the soliton is also
considered. Under certain conditions, existence of fermionic zero
modes results in fractional fermion number being induced on the classical
solution \cite{Jackiw1976,Jackiw1977}. Such systems can not relax
to trivial vacuum in isolation \cite{Sahu2004} due to Qunatum Mechanics.
This possibility was first emphasised in \cite{Vega1978} and its
consequences to possible particle like states in $SO(10)$ Grand Unified
Theory were studied in \cite{Stern1983}. Fractional fermion number
phenomenon is also of importance in condensed matter systems like
conducting polymers \cite{JackiwShrieffer}. 

In this paper, we study toroidal configurations of the Nielsen-Olesen
string. From classical arguments, this object can be shown to be unstable
with respect to shrinking under its own tension. However, it exists
as an extremum of the action in $3+1$ dimensions and carries finite
total energy, and can be of fundamental significance to cosmology.
Here we have studied the interaction of this object with a majorana
fermion field and have shown the existence of fermionic zero modes.
Significance of such solutions to cosmology was studied in \cite{steandyaj}
and more recently in \cite{Davis:1997bu,Davis:1999wq,DavKibetal}.
We use the same fermionic coupling as was first studied by Jackiw
and Rossi \cite{Jackiw1981}, wherein the mass of the fermion is derived
entirely from spontaneous symmetry breaking. Our main result about
relation between winding number and number of zero modes is the same
as \cite{Jackiw1981}, although some of the details are different.
Explicit arguments for the stability of objects with fractional fermion
number were spelt out in \cite{Sahu2004}, which will essentially
apply in the present case as well. 

Several standard caveats apply to the present work. We treat the fermions
as a qunatum perturbation to a classical background and ignore the
back reaction of the fermions to the string. While we study the highly
symmetric object, the torus, the result about zero modes should apply
in the general situation subject to some modifications, which however
should not modify the main results regarding induced stability. In
particular the {}``zero-energy'' solutions will no longer be so
on a generic closed loop geometry, however the modes, if singleton,
more generally in odd number should remain so as long as the essential
topological aspects of the boundary conditions far from the string
are not modified. Finally we also assume the metastability of the
loop to ensure its existence over time scales long enough to treat
zero-modes as occuring on essentially static background. 

We solve the Majorana-Dirac equation in toroidal coordinates, and
for this purpose begin with a review of the same in sec. \ref{sec:Toroidal-Coordinates}.
In sec. \ref{sec:Formulation-of-the} the equations are formulated.
The asymptotic form of the fermionic wave function is studied in sec.
\ref{sec:Asymptotic-analysis}. In sec. \ref{sec:Counting-the-number},
the behaviour near the core of the torus, which determines the number
of zero-energy solutions is studied, and is found to reveal that the
number is the same as for the uncompactified Nielsen-Olesen string.
Sec. \ref{sec:Conclusion} is devoted to summary and conclusion.

\section{Toroidal Coordinates\label{sec:Toroidal-Coordinates}}

The coordinate transformations from the cartesian to the toroidal
co-ordiantes \cite{MFII} are given by the following relations\begin{equation}
x=\frac{a\sinh v\cos\varphi}{\cosh v-\cos u}\end{equation}
\begin{equation}
y=\frac{a\sinh v\sin\varphi}{\cosh v-\cos u}\end{equation}
\begin{equation}
z=\frac{a\sin u}{\cosh v-\cos u}\end{equation}
where $v$ ranges from $0$ to $\infty$, $u$ ranges from $0$ to
$2\pi$ and $\varphi$ ranges from $0$ to $2\pi$. 
The parameter $a$ sets the size of the family of torii given by $v=$constant.
The variable $\varphi$ parameterises the length of the loop while
$u$ winds around any segment of the loop given by $\varphi=$constant.
The coordinates have the property that as $v$ tends to infinity,
we approach the core of the loop. Spatial infinity is approached when
$u$ and $v$ simultaneously approach zero. 

In the following it is convenient to introduce $\xi=(u+iv)/2$ and
calculate the metric elements $h_{v}=|\partial\vec{r}/\partial v|$
both in terms of $u$, $v$ and $\xi$, $\bar{\xi}$\begin{equation}
h_{u}=\frac{a}{\cosh v-\cos u}=\frac{a}{2\sin\xi\sin\bar{\xi}}\end{equation}
\begin{equation}
h_{v}=\frac{a}{\cosh v-\cos u}=\frac{a}{2\sin\xi\sin\bar{\xi}}\end{equation}
\begin{equation}
h_{\varphi}=\frac{a\sinh v}{\cosh v-\cos u}=\frac{a\sinh v}{2\sin\xi\sin\bar{\xi}}\end{equation}
 The expression for the gradient takes the form, \begin{equation}
\nabla=\hat{v}\frac{1}{h_{v}}\frac{\partial}{\partial v}+\hat{u}\frac{1}{h_{u}}\frac{\partial}{\partial u}+\hat{\varphi}\frac{1}{h_{\varphi}}\frac{\partial}{\partial\varphi}\end{equation}

\section{Formulation of the Equation\label{sec:Formulation-of-the}}

The Abelian Higgs model with gauge field $A_{\mu}$ and a charged
scalar field $\phi$ has the Lagrangian \begin{equation}
\mathcal{L}=(D_{\mu}\phi)^{*}(D^{\mu}\phi)-\frac{1}{4}F_{\mu\nu}F^{\mu\nu}-\frac{\lambda}{4}(|\phi|^{2}-\eta^{2})^{2}\end{equation}
 with $F_{\mu\nu}=\partial_{\mu}A_{\nu}-\partial_{\nu}A_{\mu}$ and
the covariant derivative \begin{equation}
D_{\mu}=(\frac{\partial}{\partial x^{\mu}}-iqh_{\mu}A_{\mu})\end{equation}
 and the scalar field $\phi$ is taken to have charge $q=e$. This
Lagrangian can be extended for the purpose of studying to zero modes
\cite{Jackiw1981}\cite{Davis-IntJ99}, to include majorana fermions
of charge $q=\frac{1}{2}e$, \begin{equation}
\mathcal{L}_{fermion}=\bar{\psi}i\sigma^{\mu}D_{\mu}\psi-\frac{1}{2}[ig_{Y}\bar{\psi}\phi\psi^{c}+(h.c.)]\end{equation}
 where $\sigma^{\mu}=(-I,\sigma^{i})$, $I$ being the $2\times2$
identity matrix; $\psi^{c}=i\sigma^{2}\psi^{*}$ is the charge conjugate
of $\psi$ , and $g_{Y}$ denotes the Yukawa coupling. 

Since $\psi$ will be reserved for later use, we begin by writing
the field equation in terms of the variable $\tilde{\psi}$\begin{equation}
\sigma^{\mu}D_{\mu}\tilde{\psi}-g_{Y}\phi\tilde{\psi}^{c}=0\end{equation}
 We begin by writing the equations of motion of the fermion in usual
cylindrical polar coordinates, $(r,\varphi,z)$ with the string loop
of radius $a$ and with cross-section $\ll a^{2}$laid out symmetrically
around the origin in the $z=0$ plane. The $\varphi$ coordinate remains
the same upon transforming to the toroidal coordinates. In $2$-component
notation, \begin{equation}
\left[\begin{array}{cc}
-e^{i\varphi}[D_{r}+\frac{i}{r}D_{\varphi}] & D_{z}+D_{t}\\
D_{z}-D_{t} & e^{-i\varphi}[D_{r}-\frac{i}{r}D_{\varphi}]\end{array}\right]\left[\begin{array}{c}
\tilde{\psi}_{1}\\
\tilde{\psi}_{2}\end{array}\right]=g_{Y}\phi\left[\begin{array}{c}
\tilde{\psi}_{1}^{*}\\
\tilde{\psi}_{2}^{*}\end{array}\right]\end{equation}
 Since we are looking for zero modes i.e. time independant solutions
we take the backgorund fields to possess the ansatz $A_{0}=0$. Further,
the lowest energy and therefore the most symmetric background solution
can be assumed $\varphi$ independent, and we choose $A_{\varphi}=0$.
However $\varphi$ explicitly appears in the equations for $\tilde{\psi}$
and factoring out this dependence requires us to introduce the ansatz
$\tilde{\psi}_{1}=e^{-i\varphi/2}\psi_{1}$ and $\tilde{\psi}_{2}=e^{i\varphi/2}\psi_{2}$.
This amounts to anti-periodic boundary condition appropriate to a
fermion as we traverse the length of the loop. Then the equations
obeyed by $\psi_{1}$ and $\psi_{2}$ are \begin{equation}
\left[\begin{array}{cc}
-[D_{r}+\frac{1}{2r}] & D_{z}\\
D_{z} & [D_{r}+\frac{1}{2r}]\end{array}\right]\left[\begin{array}{c}
\psi_{1}\\
\psi_{2}\end{array}\right]=g_{Y}\phi\left[\begin{array}{c}
\psi_{1}^{*}\\
\psi_{2}^{*}\end{array}\right]\end{equation}
 Thus the problem of solving fermionic equations is restricted essentially
to the half plane of the cylindrical polar coordinates, $r\in[0,\infty)$
and $z\in(-\infty,\infty)$. Substituting $\psi_{1}=i\psi_{2}=i\psi$
reduces the two equations to the complex equation \begin{equation}
[D_{r}+iD_{z}]\psi+\frac{1}{2r}\psi=g_{Y}\phi\psi^{*}\end{equation}

We now transform to the toroidal coordinates. Since $\partial_{r}$
and $A_{r}$ transform identically. Thus, transforming to toroidal
coordinates the equation looks as follows, \begin{equation}
[\sin^{2}\bar{\xi}(D_{u}+iD_{v})+i\frac{\sin\xi\sin\bar{\xi}}{2\sinh v}]\psi=\frac{\phi}{2i}ag_{Y}\psi^{*}\end{equation}
 A useful substituion now is $\psi=f(u,v)\sin\xi$, which leads to
the equation for $f$, \begin{equation}
[\sin\xi\sin\bar{\xi}(D_{u}+iD_{v})+i\frac{\sin^{2}\xi}{2\sinh v}]f=\frac{\phi}{2i}ag_{Y}f^{*}\label{toroidal}\end{equation}
 We work in the vacuum sector of winding number $n$, i.e., given
any segement of the loop, the scalar field $\phi$ changes phase by
$2\pi n$ around it. In toroidal coordinates this amounts a dependence
$e^{inu}$. While topologically this is not distinct from trivial
vacuum, it has restricted topologically stability against breaking
of any segment of the loop. Only the shrinking of the loop as a whole
can continuously connect it to the trivial vacuum. Thus if latter
deformation is forbidden the configuration becomes stable. As a direct
generalisation of the Nielsen-Olesen string, the background field
configuration is taken to have the ansatz \begin{equation}
\phi=ik(u,v)\eta e^{inu}\end{equation}
\begin{equation}
A_{\mu}=-n\frac{g(u,v)}{ae}\delta_{u}^{\mu}\sin\xi\sin\bar{\xi}\end{equation}
 where $g(u,v)$ and $k(u,v)$ are real functions whose behaviour
is $g,\, k\rightarrow0$ near the loop, i.e., as $v\rightarrow\infty$
so that the solution is regular in core of the loop, and $g(u,v),\, k(u,v)\rightarrow1$
at spatial infinity given by the simultaneous limit $u,\, v\rightarrow0$.
Note that $\sin\xi\sin\bar{\xi}$reproduces the well known behaviour
$1/r$ for usual infinitely long string in cylindrical coordinates
and becomes pure gauge far from the core of the string. We denote
$g_{Y}\eta=m$ where $m$ is the mass of the free fermions far from
the string. After substituting above scalar and gauge ansatz, eq.
(\ref{toroidal}) takes the form \begin{equation}
[\sin\xi\sin\bar{\xi}(\frac{\partial}{\partial\bar{\xi}}-i\frac{n}{2}g)+i\frac{\sin^{2}\xi}{2\sinh v}]f=\frac{1}{2}akme^{inu}f^{*}\end{equation}
 Using the technique of Jackiw and Rossi, we try the following ansatz
for $f$. 

\begin{equation}
f=Xe^{il(\xi+\bar{\xi})}+Y^{*}e^{i(n-l)(\xi+\bar{\xi})}\end{equation}
 and equating coefficients of $e^{ilu}$ and $e^{i(n-l)u}$ we get
two separate equations.\begin{equation}
[\sin\xi\sin\bar{\xi}(\frac{\partial}{\partial\bar{\xi}}-i(\frac{n}{2}g-l))+i\frac{\sin^{2}\xi}{2\sinh v}]X=\frac{1}{2}akmY\label{main1}\end{equation}
\begin{equation}
[\sin\xi\sin\bar{\xi}(\frac{\partial}{\partial\bar{\xi}}-i(\frac{n}{2}g-(n-l)))+i\frac{\sin^{2}\xi}{2\sinh v}]Y^{*}=\frac{1}{2}akmX^{*}\label{cc}\end{equation}
 taking complex conjugate of the Eq. ($\ref{cc}$),\begin{equation}
[\sin\xi\sin\bar{\xi}(\frac{\partial}{\partial\xi}+i(\frac{n}{2}g-(n-l)))-i\frac{\sin^{2}\bar{\xi}}{2\sinh v}]Y=\frac{1}{2}akmX\label{main2}\end{equation}

\section{Asymptotic analysis\label{sec:Asymptotic-analysis}}

In the asymptotic limit, as mentioned above, $g$ and $k$ can be
approximated by $1$. So in the asymptotic limit the equations ($\ref{main1}$)
and ($\ref{main2}$) respectively become,\begin{equation}
[\xi\bar{\xi}(\frac{\partial}{\partial\bar{\xi}}-ip)-\frac{1}{2}\frac{\xi^{2}}{(\xi-\bar{\xi})}]X=\frac{1}{2}amY\end{equation}
\begin{equation}
[\xi\bar{\xi}(\frac{\partial}{\partial\xi}-ip)+\frac{1}{2}\frac{\bar{\xi}^{2}}{(\xi-\bar{\xi})}]Y=\frac{1}{2}amX\end{equation}
 where $p=(\frac{n}{2}-l)$. We substitute, $X=A\sqrt{\frac{4i\xi\bar{\xi}}{\xi-\bar{\xi}}}$
similarly $Y=B\sqrt{\frac{4i\xi\bar{\xi}}{\xi-\bar{\xi}}}$. The equations
are simplified to,\begin{equation}
\xi\bar{\xi}(\frac{\partial}{\partial\bar{\xi}}-ip)A=\frac{1}{2}amB\label{2main1}\end{equation}
\begin{equation}
\xi\bar{\xi}(\frac{\partial}{\partial\xi}-ip)B=\frac{1}{2}amA\label{2main2}\end{equation}
 combining Eq. ($\ref{2main1}$) and ($\ref{2main2}$) we get,\begin{equation}
[\xi\bar{\xi}(\frac{\partial}{\partial\xi}-ip)\xi\bar{\xi}(\frac{\partial}{\partial\bar{\xi}}-ip)]A=(ma/2)^{2}A\label{main}\end{equation}
 Substituting $\xi=t^{-1}e^{i\theta}$, Eq. ($\ref{main}$) is simplified
to\begin{equation}
[\frac{\partial^{2}}{\partial t^{2}}+(\frac{2ip}{t^{2}}-\frac{1}{t})\frac{\partial}{\partial t}-(\frac{p^{2}}{t^{4}}+\frac{2ip}{t^{3}}+(\frac{ma}{2})^{2})+\frac{1}{t^{2}}(\frac{\partial^{2}}{\partial\theta^{2}}+2i\frac{\partial}{\partial\theta})]A=0\end{equation}
 This, in asymptotic limit, i.e. as $|\xi|\rightarrow0$, i.e. as
$t\rightarrow\infty$, becomes\begin{equation}
[\frac{\partial^{2}}{\partial t^{2}}-\frac{1}{t}\frac{\partial}{\partial t}-(\frac{ma}{2})^{2}]A=0\label{final}\end{equation}
 this second order differential equation can be solved to yield an
exponentially converging solution, whose asymptotic behaviour is $\sim e^{-mat/2}$
. Incorporating $\sin\xi$ factor, the asymptotic behaviour of the
fermionic wave-function $\psi\sim(e^{-mat/2})/t$ making it normalisable.

\section{Counting the number of solutions\label{sec:Counting-the-number}}

To count the total number of fermion zero modes present on soliton
in $n$ vortex sector, we observe the $v$ dependance of the solution
near the loop i.e. as $v\rightarrow\infty$. As mentioned above $g$
and $k$ both tend to zero as we approach the loop. So substituting
$g=0=k$ the equations ($\ref{main1}$) and ($\ref{main2}$), near
the loop, respectively become,\begin{equation}
(\frac{d}{dv}+l)X=0\end{equation}
\begin{equation}
(\frac{d}{dv}+(n-l))Y=0\end{equation}
So, $X\sim e^{-lv}$and $Y\sim e^{-(n-l)v}$. And so the behaviour
of $\psi$ near the loop is,\begin{equation}
\psi=\sin\xi(Xe^{ilu}+Y^{*}e^{i(n-l)u})=\sin\xi(C_{1}e^{2il\xi}+C_{2}e^{2i(n-l)\xi})\label{near}\end{equation}
 In the limit $v\rightarrow\infty$ the right hand side of eq. ($\ref{near}$)
is dominated by the terms\begin{equation}
\psi\longrightarrow\frac{1}{2i}(C_{1}e^{i2(l-\frac{1}{2})\xi}+C_{2}e^{i2[(n-1)-(l-\frac{1}{2})]\xi})\end{equation}
 Recalling $\xi=(u+iv)/2$ and denoting $l-\frac{1}{2}$ by $l^{\prime}$,
and requiring $\psi$ to remain finite near the loop i.e. as $v\rightarrow\infty$,
we need,\begin{equation}
0\leq l^{\prime}\leq(n-1)\end{equation}
 This gives us total of $n$ complex normalisable solutions, the same
result as \cite{Jackiw1981} for the infinitely long string. It should
be noted that we have the $\varphi$ dependence $e^{\pm i\varphi}$.
If the length parameter along the loop is denoted $\tilde{z}$, this
can be written as $e^{\pm i\tilde{z}/2\pi a}$. This explicit dependence
on $\tilde{z}$ disappears in the limit $a\rightarrow\infty$and we
recover the translation invariant ansatz for the zero modes utilised
in \cite{Jackiw1981}. Taking $l^{\prime}$ to be integer (rather
than half-integer) gives larger number of solutions and makes the
latter single valued as functions of $u$, which also accords with
the treaatment for infinitely long string. So the compactification
of Nielsen-Olesen string has not altered the number of zero modes
it carries.

\section{Conclusion\label{sec:Conclusion}}

We have proved the existence of $|n|$ fermionic zero modes on a static
toroidal string with topological winding $n$. Unlike the non-compact
Nielsen-Olesen strings which are infinitely long and often treated
as essentially $2+1$ dimensional solitons, toroidal strings are genuinely
$3+1$ dimensional configurations of finite energy. So the existence
of the latter and the existence of related zero modes are very important
from the point of view of cosmology. Our result shows that the toroidal
geopmetry supports the same number of zero modes as the infinitely
long string and reassures us that the unbounded string can be recovered
as a limiting case of the toroidal configurations considered here. 

The boundary condition implied by the behaviour $e^{\pm i\varphi/2}$
with azimuthal angle $\varphi$ shows that for small loops, when the
loop is indistinguishable form a particle, its wave fundtion obeys
the same boundary ocnditions as an elementary fermion of spin $1/2$.
Physically such states should be discovered as heavy fermions of spin
$1/2$. Further, the occurence of zero modes would imply, just as
in the case of unbounded string, that the loop acquires fermionic
charge $|n|/2$. If this charge is half-integral, it would be impossible
for the loop to disintegrate in isolation without conflicting with
Quantum Mechanics. The arguments detailed in \cite{Sahu2004} apply
without significant modification. 

When these considerations are further applied to the collective dynamics
of the string, new situations need to be addressed. Consider a loop
of large radius which folds and begins to cross itself. In the absence
of experimental evidence and absence of conclusive theoretical calculation
two possibilites are usually considered, one where the two colliding
segments pass through and the other where they inter-commute, producing
two smaller loops. Since the winding number of the two child strings
would be the same as the parent string the number zero modes on each
of the child strigns would be the same as the parent string. If therefore
the parent string had half-integer fermion number, the final state
would have integer fermion number. To avoid conflict with quantum
mechanical principles we must insist that the inter-commuting process
cannot occur for the strings with odd number of zero-modes. 

In \cite{Sahu2004} it was explicitly shown that a single non-compact
string cannot decay in isolation even if metastable. However no conclusion
could be reached about formation of loops formed by self-intersection
of a non-compact string. With the results of the present paper we
can conclude that formation of loop a by such a process is also forbidden
for non-compact strings with odd number of zero-modes, for the same
reason as in preceeding paragraph. 

Loops stabilised by quantum mechanical considerations would be extremely
important to Cosmology, where such loops can constitute Cold Dark
Matter \cite{KolTur}\cite{ParamWMAP}. We may assume that the process
of shrinking of the loop under its own tension can continue till some
small radius is reached, presumably of the order of the Compton wavelength
of the fermions. Provided that fermions are much lighter, such a length
would be large compared to the cross-section of the string characterised
by gauge boson and scalar masses. Such a state would then be indistinguishable
for classical purposes from a fundamental particle. While all the
mutually interacting particles would decay into the lightest available
particle state subject to conserved quantum numbers, heavy states
such as stabilised string loops would persist and serve as Dark Matter.
Conversely, unified theories implying unacceptable abundance of such
stabilised loops would be ruled out by such considerations.

\section*{Acknowledgement}

This work is supported by a grant from Department of Science and Technology.

\end{document}